\documentclass[10pt,aps,prd,notitlepage,twocolumn,superscriptaddress,nofootinbib,amsmath,amssymb]{revtex4-1}
\usepackage{graphicx}
\usepackage{cmap}
\usepackage[utf8]{inputenc}
\usepackage[T1]{fontenc}
\usepackage{xcolor}

\usepackage{orcidlink}

\def\Order#1{{\cal O}\left(#1\right)}

\begin{document}


\title{Time Evolution of Black Hole Perturbations in Quadratic Gravity}

\author{Roman A. Konoplya}
\email{roman.konoplya@gmail.com}
\affiliation{Research Centre for Theoretical Physics and Astrophysics,
Institute of Physics, Silesian University in Opava,
Bezručovo náměstí 13, CZ-74601 Opava, Czech Republic}

\author{Andrea Spina}
\email{Andrea.spina@phd.unict.it}
\affiliation{Department of physics and astronomy, Università di Catania, via S. Sofia 64, I-95123,Catania, Italy}
\affiliation{INFN, Sezione di Catania,  via S. Sofia 64, I-95123,Catania, Italy}
\affiliation{Research Centre for Theoretical Physics and Astrophysics,
Institute of Physics, Silesian University in Opava,
Bezručovo náměstí 13, CZ-74601 Opava, Czech Republic}

\author{Alexander Zhidenko}
\email{olexandr.zhydenko@ufabc.edu.br}
\affiliation{Research Centre for Theoretical Physics and Astrophysics,
Institute of Physics, Silesian University in Opava,
Bezručovo náměstí 13, CZ-74601 Opava, Czech Republic}
\affiliation{Centro de Matemática, Computação e Cognição (CMCC), Universidade Federal do ABC (UFABC),
Rua Abolição, CEP: 09210-180, Santo André, SP, Brazil}
 
\begin{abstract}
We study the full time-domain evolution of gravitational perturbations in black hole spacetimes arising in Einstein–Weyl gravity, a renormalizable extension of general relativity containing quadratic curvature corrections. We analyze both Schwarzschild and non-Schwarzschild solutions, focusing on monopole and higher multipole perturbations. Using semi-analytical methods based on the Rezzolla–Zhidenko parametrization for approximation of the black hole spacetime and time-domain integration for analysis of evolution of perturbations, we study the late-time behavior of gravitational perturbations. Our results show that the ringdown phase is followed by universal slowly decaying oscillatory tails with the envelope \( \psi \propto t^{-5/6} \). We also demonstrate the breakdown of the eikonal correspondence between quasinormal modes and unstable null geodesics, highlighting limitations of the WKB method in this context. Our analysis confirms the range of (in)stability of black holes in Einstein–Weyl gravity found in recent publications. 
\end{abstract}

\maketitle

\section{Introduction}

The study of black hole perturbations offers a powerful lens through which to probe the nature of gravity and the underlying structure of spacetime. Quasinormal modes (QNMs) \cite{Nollert:1999ji,Konoplya:2011qq,Kokkotas:1999bd}, which characterize the damped oscillations of perturbed black holes, serve as fingerprints of the black hole geometry and the gravitational theory in question. With the advent of gravitational wave astronomy \cite{LIGOScientific:2016aoc,LIGOScientific:2017vwq,LIGOScientific:2020zkf,LISA:2022yao,LISA:2022kgy}, QNMs have acquired central importance as tools for testing general relativity and its alternatives in the strong-field regime. Among the various modified theories of gravity, Einstein-Weyl gravity --- a higher-derivative extension of general relativity --- has drawn attention due to its ability to remain perturbatively renormalizable while introducing novel features into the structure of black hole solutions. In this context, understanding the time evolution of gravitational perturbations and their spectral signatures becomes crucial for assessing the stability and observational viability of such quantum-corrected black holes.

Black holes in Einstein-Weyl gravity exhibit a number of distinctive features. Most notably, there exist two asymptotically flat black hole solutions: one corresponding to the Schwarzschild metric and the other to a non-Schwarzschild solution, which arises for specific values of the coupling parameter $\alpha$. The non-Schwarzschild solution is particularly interesting in that, as $\alpha$ approaches its maximal value — still permitting the existence of a horizon — the black hole's mass tends to zero, while the radius remains finite. Such ultra-light black holes are unlikely to form through stellar collapse,  it is therefore reasonable to expect astrophysical processes to produce Schwarzschild black holes, making the crossing point relevant only in the final stages of black hole evaporation. The Schwarzschild solution branch intersects with the Non-Schwarzschild one \cite{Silveravalle:2022wij} at a specific mass, which can range from that of an asteroid to the Planck mass \cite{Bonanno:2024fcv}, based only on the value of a constrained free parameter obtained with torsion test \cite{Giacchini:2016nta}. 

While perturbations and quasinormal modes of black holes in Einstein-Weyl gravity have been thoroughly explored for test fields in \cite{Zinhailo:2018ska,Zinhailo:2019rwd,Konoplya:2022iyn}, studies of gravitational perturbations have only recently begun to emerge \cite{Antoniou:2024jku, Held:2022abx}. The spectrum of test fields in this theory has shown distinctive behavior: for instance, when the coupling $\alpha$ is near its minimal value and slightly increased, the fundamental mode remains nearly unchanged, whereas higher overtones deviate progressively more strongly \cite{Konoplya:2022iyn}. This pronounced sensitivity of overtones is attributed to static deformations of the near-horizon geometry induced by the coupling. This effect, sometimes referred to as “the sound of the event horizon” \cite{Konoplya:2023hqb}, has recently attracted growing interest and has been investigated for various black hole models \cite{Konoplya:2025hgp,Zinhailo:2024kbq,Konoplya:2024lch,Konoplya:2023kem,Stashko:2024wuq,Konoplya:2023ahd,Stuchlik:2025ezz,Lutfuoglu:2025ljm}.

Another notable feature of Einstein-Weyl gravity is the emergence of an effectively massive channel in the spectrum of gravitational perturbations. This additional degree of freedom corresponds, at the quantum level, to a massive graviton, introducing new dynamical behavior absent in general relativity.
The presence of an effective mass significantly alters the spectrum of quasinormal modes and the asymptotic behavior of perturbations (see, for example, \cite{Lutfuoglu:2025hjy,Zinhailo:2024jzt,Dubinsky:2024jqi,Bolokhov:2023bwm,Bolokhov:2023ruj,Malik:2024cgb,Dubinsky:2024hmn,Konoplya:2024wds,Davlataliev:2024mjl,Lutfuoglu:2025hwh} for recent developments). One particularly interesting phenomenon is the emergence of quasinormal modes with arbitrarily small damping rates for massive fields. These so-called \textit{quasi-resonances} appear at critical values of the field mass, leading to exceptionally long-lived modes in the spectrum \cite{Ohashi:2004wr,Konoplya:2004wg,Konoplya:2006br}. However, the existence of quasi-resonances is not a universal feature of massive perturbations. Notable exceptions include massive vector fields \cite{Konoplya:2005hr}, certain brane-world scenarios \cite{Zinhailo:2024jzt}, and perturbations in asymptotically de Sitter spacetimes \cite{Konoplya:2004wg,Lutfuoglu:2025hjy}. Therefore, the behavior of massive or effectively massive field perturbations must be carefully analyzed on a case-by-case basis.

Perturbations of massive fields also have promising observational potential. Following the ringdown phase, these perturbations can give rise to slowly decaying oscillatory tails, which contribute to ultra-long-wavelength radiation \cite{Konoplya:2023fmh}. Such signals are of particular interest in the context of recent observations by the Pulsar Timing Array experiment \cite{NANOGrav:2023hvm,NANOGrav:2023hvm}.

While the dominant quasinormal modes in Einstein-Weyl gravity have been explored — though not exhaustively — in recent works \cite{Antoniou:2024jku,Held:2022abx,Zinhailo:2018ska,Zinhailo:2019rwd,Konoplya:2022iyn}, a comprehensive analysis of the full time-domain evolution of gravitational perturbations is still lacking. In particular, no study to date has captured the slowly decaying oscillatory tails that follow the ringdown phase — features crucial for assessing both the stability of black hole solutions and the detectability of hypothetical associated signals, such as those targeted by primordial black holes or Pulsar Timing Array experiments. In this paper, we aim to fill this gap by examining the complete time evolution of gravitational perturbations for both known branches of black hole solutions in Einstein-Weyl gravity.

We show that gravitational perturbations of Schwarzschild and non-Schwarzschild black holes in Einstein–Weyl gravity exhibit distinctive dynamical signatures, including long-lived oscillatory tails, instabilities, and deviations from the classical null geodesic correspondence. By employing numerical method of integration of the wave equations in time-domain, we systematically study the evolution of monopole and higher-multipole perturbations across different values of the dimensionless parameter, extracting quasinormal frequencies and characterizing asymptotic decay rates.

The paper is organized as follows. In Sec.~\ref{sec:weyl}, we introduce the Einstein–Weyl theory as a particular limit of quadratic gravity and discuss the black hole solutions it admits. We also present a semi-analytical approximation based on the Rezzolla–Zhidenko parametrization, which facilitates accurate numerical computations. In Sec.~\ref{sec:perturbations}, we analyze gravitational perturbations for various multipole numbers, focusing on both Schwarzschild and non-Schwarzschild branches. The regime of large multipole numbers is considered in Sec.~\ref{sec:eikonal}, where we discuss the applicability of the WKB method and the null geodesic correspondence. Finally, in Sec.~\ref{sec:conclusion}, we summarize our main results and discuss their broader implications.

\section{Quadratic gravity}\label{sec:weyl}

We begin with the gravitational action extended by quadratic curvature terms, which provides a natural generalization of Einstein’s theory \cite{Stelle:1977ry}:
\begin{equation}
    \mathcal{S}=\int {\rm d}^4 x \sqrt{-g} \left( R-\alpha C_{\mu \nu \rho \sigma} C^{\mu \nu \rho \sigma}+\beta R^2 \right)\,,
\label{eq:action}
\end{equation}
where $R$ is the Ricci scalar and $C_{\mu \nu \rho \sigma}$ is the Weyl tensor.


It is well known \cite{Hindawi:1995an} that this action introduces additional dynamical degrees of freedom associated with the couplings $\alpha$ and $\beta$. Specifically, in the Einstein frame, the theory describes a massive scalar particle and a massive spin-2 tensor particle with mass ($\mu^2=1/2\alpha$), the latter possessing negative energy states, which leads to certain peculiarities in the physical interpretation of the theory. It is also noteworthy that all solutions of the full quadratic theory have vanishing Ricci scalar. In particular, any static and asymptotically flat black hole solution in the full theory also satisfies the field equations with $\beta=0$, corresponding to Einstein-Weyl (EW) gravity \cite{Lu:2015psa, Lu:2015cqa, Bonanno:2019rsq}. From this point onward, we will focus exclusively on the EW theory.

\subsection{Field equation and Background solutions}
We derive the equations of motion by varying the action (\ref{eq:action}), leading to \cite{Lu:2015psa}
\begin{equation}\label{eomham}
\left(R_{\mu\nu}-\frac{1}{2}R\,g_{\mu\nu}\right)-4\alpha\left(\nabla^\rho\nabla^\sigma+\frac{1}{2}R^{\rho\sigma}\right)C_{\mu\rho\nu\sigma}=0.
\end{equation}
Our objective is to reduce the field equations to a second-order form. To this end, we introduce an auxiliary field \( f_{\mu\nu} \) into the action, following a standard approach in the literature \cite{Antoniou:2024jku, Held:2022abx}. By varying the modified action with respect to \( g_{\mu\nu} \) and \( f_{\mu\nu} \), we obtain the following equations of motion. For \( g_{\mu\nu} \), we have:
\begin{equation}
     \mathcal{E}^{(g)}_{\mu\nu}\equiv G_{\mu\nu}+ \mu^2 \left(f_{\mu\nu}-f g_{\mu\nu}\right)=0\, ,
\label{eq:field_g}
\end{equation}
and consequently, for $f_{\mu\nu}$ \cite{Antoniou:2024jku} 
\begin{equation}
\begin{split}
    &
    \mathcal{E}^{(f)}_{\mu\nu}\equiv
    G_{\mu\nu}+G_{(\mu}{}^\rho f_{\nu)\rho}-g_{\mu\nu}G^{\rho\sigma}f_{\rho\sigma}-fR_{\mu\nu}
    \\
    &
    +R f_{\mu\nu}+\Box f_{\mu\nu}+(\nabla_\mu\nabla_\nu-g_{\mu\nu})f
    -2\nabla_\rho\nabla_{(\mu}f_{\nu)}{}^\rho\\
    &
    +g_{\mu\nu}\nabla_\rho\nabla_\sigma f^{\rho\sigma}+\mu^2\big[\big(f^2-f_{\rho\sigma}f^{\rho\sigma}\big)/2\\
    &
    +2\big(f_\mu{}^\rho f_{\nu\rho}-f\,f_{\mu\nu}\big)\big]=0.
 \end{split}
\label{eq:field_f}
\end{equation}
We note that these equations resemble those of classical General Relativity, with additional terms that vanish in the limit when $\alpha$ goes to zero, recovering Einstein’s equations.

Now, we turn to black hole solutions. We start with the general metric for a static, spherically symmetric spacetime:
\begin{equation}
    d s^2=-A(r) d t^{2} +\frac{1}{B(r)} d r^{2}+ r^{2}d\Omega^2.
\label{eq:metric}
\end{equation}
Substituting this metric into the equations of motion, we obtain the differential equations for the metric functions $A$ and $B$
\begin{equation}
\begin{split}
    &
    2 r^2 A^2 B (r A'-2 A) B''-B' \big[4 r A^3-r A B (r^2 A'^2\\
    &
    +2 r A A'+4 A^2)\big]-4 A^2 B \big[\mu ^2 r^3 A'+A (\mu ^2 r^2+2)\big]\\
    &
    -B^2 (r^3 A'^3-3 r^2 A A'^2-8 A^3)+3 r^2 A^3 B'^2\\
    &
    +4 \mu ^2 r^2 A^3=0\, .
\end{split}
\label{eq:bg_2}
\end{equation}
To find solutions to these equations, we employ approximations, particularly by analyzing the behavior at large distances and near the horizon.

At large distances, we impose asymptotic flatness, allowing us to use the weak-field limit, where the metric is a perturbation of Minkowski spacetime. Expanding linearly in $\epsilon$  
 $A(r)=1+\epsilon\,V(r)$ \cite{Lu:2015psa}, and similarly for $B$,  we obtain the linearized solution \cite{Bonanno:2019rsq,Silveravalle:2023lnl}
\begin{equation}\begin{split} \label{hflinear}
A(r)=\, &1-\frac{2\,M}{r}+2S^-_2 \frac{e^{-\mu\, r}}{r}, \\
B(r)=\, &1-\frac{2\,M}{r}+S^-_2 \frac{e^{-\mu\, r}}{r}(1+\mu\, r) ,
\end{split}\end{equation}
where we have two free parameters, $M$ which is the ADM mass (expressed in Planck units which we will use throughout this work, so that $\mu$ and $\alpha$ are proportional to the Planck mass and length \cite{Silveravalle:2023lnl,Held:2022abx}) and $S^-_2$ that we can define as Yukawa ``charge'' due to the form of the correction in the approximation.

Near the horizon, we employ a generalized Frobenius series expansion for the metric functions:
\[
\begin{split}
A(r)&=\left(r-r_0\right)^t\!\left[\displaystyle\sum_{n=0}^N h_{t+\frac{n}{\Delta}}\left(r-r_0\right)^{\frac{n}{\Delta}}\!+\!\Order{r-r_0}^{\frac{N\!+\!1}{\Delta}}\!\right]\!,\\
B(r)&=\left(r-r_0\right)^s\!\left[\displaystyle\sum_{n=0}^N f_{s+\frac{n}{\Delta}}\left(r-r_0\right)^{\frac{n}{\Delta}}\!+\!\Order{r-r_0}^{\frac{N\!+\!1}{\Delta}}\!\right]\!,
\end{split}
\]
(it is possible to classify different solutions with values of $s,t$ \cite{Lu:2015psa,Silveravalle:2022wij,perkinsthesis}). We are interested in the study of black holes and as done in \cite{Antoniou:2024jku} we distinguish two families: one characterized by ${R}_{\mu\nu}=0$ corresponding to Schwarzschild black holes, and another with ${R}=0$ which differs due to the presence of $f_{\mu\nu}$ also in the background leading to what we refer to as Non-Schwarzschild black holes.

\subsection{Semi-analytical approximation}\label{approx}

For numerical computations, it is often more efficient to adopt the general parametrization for asymptotically flat, spherically symmetric black holes in arbitrary metric theories of gravity, known as the Rezzolla–Zhidenko parametrization~\cite{Rezzolla:2014mua}. This approach is conceptually similar to the post-Newtonian parametrization, but unlike the latter, it remains valid across the entire domain outside the event horizon, extending all the way to spatial infinity.

The Rezzolla–Zhidenko framework provides a systematic and convergent expansion that captures both near-horizon and asymptotic behavior. It was later generalized to axially symmetric \cite{Konoplya:2016jvv,Younsi:2016azx} black holes and has been widely used in astrophysically motivated applications, including the analytic approximation of numerical black hole solutions~\cite{Konoplya:2021slg,Konoplya:2021qll,Kocherlakota:2020kyu,Zhang:2024rvk,Cassing:2023bpt,Konoplya:2020hyk,Li:2021mnx,Ma:2024kbu,Shashank:2021giy,Kokkotas:2017ymc,Yu:2021xen,Nampalliwar:2019iti,Toshmatov:2023anz,Ni:2016uik,Paul:2023eep}. This parametrization has proven especially useful for modeling deviations from general relativity in a theory-independent way.

The semi-analytical approximation for the metric function based on the Rezzolla-Zhidenko parametrization was  introduced in \cite{Kokkotas:2017zwt}, which reproduces the correct asymptotic behavior at infinity and at the event horizon. It has been verified against numerical solutions of eq.~\eqref{eq:bg_2} \cite{Kokkotas:2017zwt}. The metric functions are defined as
\begin{align}
    A(r) & \equiv x\,f(x)\,,\\
    \frac{A(r)}{B(r)} & \equiv h(x)^2\,,\label{eq:fh}
\end{align}
where $x = 1-r_h/r$, is a compact coordinate, with $r_h$ being the radius of the black hole event horizon. The functions $f(x)$ and $h(x)$ are parameterized as
\begin{align}
    f(x)=&1-\epsilon (1-x)-\epsilon(1-x)^2+\tilde{f}(x)(1-x)^3\,,\nonumber\\
    h(x)=&1+\tilde{h}(x)(1-x)^2\,,
\end{align}
and expressed in terms of the following continued fraction expansions 
\begin{equation}
\tilde{f}(x)=\frac{\tilde{f}_1}{1+\frac{\tilde{f}_2 x}{ 1+\frac{\tilde{f}_3 x}{1+\frac{\tilde{f}_4 x}{1+\ldots}}}}\,,
\;\tilde{h}(x)=\frac{b_1}{1+\frac{\tilde{h}_2 x}{1+\frac{\tilde{h}_3x}{1+\frac{\tilde{h}_4 x}{1+\ldots}}}}\,,
\end{equation}
where the coefficients depend on the dimensionless parameter $p=r_h\mu$.

\section{Black hole perturbations}\label{sec:perturbations}

Now we can study the linear perturbations of the black hole background, expanding:
\begin{align}
g_{\mu\nu}&=\bar{g}_{\mu\nu}+\varepsilon\, \delta g_{\mu\nu}\,,\\ 
f_{\mu\nu}&=\bar{f}_{\mu\nu}+\varepsilon\, \delta f_{\mu\nu}\,, \label{pertfg}
\end{align}
where the oversigned expression are the non perturbed metric coefficients. Then, for spherically symmetric background, we can expand perturbations into tensor spherical harmonics with the radial part separated from the angular one in a usually way \cite{Konoplya:2011qq,Antoniou:2024jku,Held:2022abx,Spina:2024npx}. 

First, we will consider various cases depending on the value of the multipole number $l$. 
\subsection{l=0}
For \( l = 0 \), when the background functions are perturbed (as in eq.~\eqref{pertfg})
and the full field equations of quadratic gravity (eqs.~\eqref{eq:field_g},~\eqref{eq:field_f}) are linearized, the resulting perturbation equations can be reduced — remarkably, as demonstrated in~\cite{Held:2022abx} — to a single wave equation of the well-known Regge--Wheeler/Zerilli form~\cite{Regge:1957td,Zerilli:1970se},
\begin{equation}
    \left(\frac{d^2}{dt^2}-\frac{d^2}{dr^2_*}\right)\psi(r,t)+V(r)\psi(r,t)=0  \label{pert}.
\end{equation}
with a general radial potential of the form
\begin{align}\label{eq:masterEq_monopole}
	&V(r)=\,
		-\mu^2\,A
		-\frac{(AB)'}{2r}
		\\&-\mu^2\,\frac{
   			24\,A^2 B\,(2\,A-r\,A')\,(2\,B + r\,B')
   		}{
   		\left(
   			-4\,\mu^2\,r\,A^2\,(3\,B-1)
   			+(AB)'\,(3\,r\,(AB)'-4\,A)
   		\right)\,r
   		}
   		\notag\\&
   		-\mu^4\,\frac{
   			288\,A^3 B^3\,(2\,A-r\,A')^2
   		}{
   		\left(
   			-4\,\mu^2\,r\,A^2\,(3\,B-1)
   			+(AB)'\,(3\,r\,(AB)'-4\,A)
   		\right)^2
   		}\;.
	\notag
\end{align}    
At infinity, the effective potential goes to  $\mu^2$,  as expected for effective \textit{massive} field perturbations.

\subsubsection{Schwarzschild black holes}
For Schwarzschild background, as said, we have $\bar{R}_{\mu\nu}=0$, so eqs.~\eqref{eq:field_g},~\eqref{eq:field_f} for the perturbations reduce to:
\begin{align}
    &
    \delta\mathcal{E}^{(g)}_{\mu\nu}=
    \delta G_{\mu\nu} + \mu^2 \delta f_{\mu\nu}=0
    \label{eq:linear-Ricciflat-eom-g}\\
    &
    \delta\mathcal{E}^{(f)}_{\mu\nu}=
    \bar{\Box} \delta f_{\mu\nu}+2\bar{R}_{\mu\sigma\nu\rho}\delta f^{\sigma\rho}-\mu^2\delta f_{\mu\nu}=0\; .
\label{eq:linear-Ricciflat-eom-f}
\end{align}
and for radial perturbation, potential of eq.~\eqref{eq:masterEq_monopole} simplifies to: 
\begin{equation}
    V_0 = f \left( \frac{2M}{r^3} + \mu^2 + \frac{24M(M - r)\mu^2 + 6r^3 (r - 4M) \mu^4}{(2M + r^3 \mu^2)^2} \right)
\end{equation}
as shown in \cite{Brito:2013wya}.

\begin{figure}[hbt!]
    \centering
    \includegraphics[width=\columnwidth]{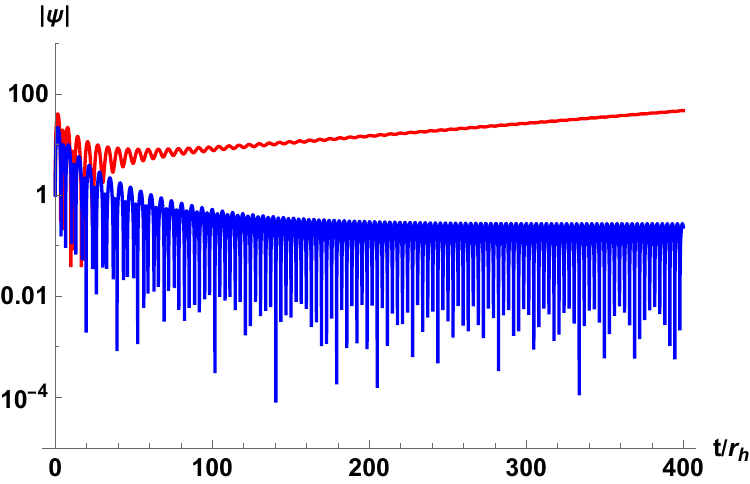}
    \caption{Semi Logarithmic plot of time evolution of massive radial peruturbation ($l=0$) for Schwarzschild background with: Blue one has $p=r_h\mu=1$, red one is the unstable case with $p=r_h\mu=0.86$.}
\label{fig1}
\end{figure}

Using these expressions, we investigated the time evolution of the perturbations (the numerical method is described in Appendix~\ref{appendix}). As the first step, we examined the stability range of the dimensionless parameter $p=r_h\mu$ for these black holes and considered the onset of the  instability of the Gregory–Laflamme type \cite{Gregory:1993vy} in light of the results reported in \cite{Held:2022abx}. Our analysis confirmed the same stability interval for $p$ ($0.87<p<1.14$), and reproduced the expected behavior associated with the Gregory–Laflamme instability, as illustrated in Fig.~\ref{fig1} for two representative values. These results are consistent with those obtained in the initial studies of the perturbation dynamics presented in \cite{Silveravalle:2023lnl, Bonanno:2024fcv} and with the general picture of time-domain evolution of the Gregory–Laflamme instability \cite{Konoplya:2008yy}.
We then proceeded to analyze the late-time asymptotic behavior for perturbations of black hole solutions in the  stability sector. The late-time tails are oscillatory and exhibit a decay of, at most, two orders of magnitude relative to the amplitude of the initial ringdown phase. This behavior contrasts with that of massless fields, where asymptotic tails typically become visible only after the signal has decayed by six or more orders of magnitude. Thus, we conclude that massive tails are not only oscillatory, but also decay significantly more slowly, as illustrated in Fig.~\ref{fig1}.

\begin{figure}[hbt!]
    \centering
    \includegraphics[width=\columnwidth]{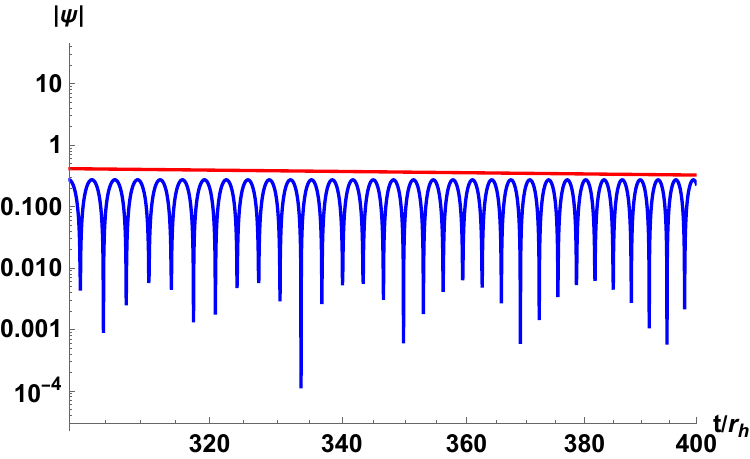}
    \caption{Logarithmic plot of the asymptotic tail part with $p=r_h\mu=1$ showed in fig.~\ref{fig1}, in red the fit of the tail with the law $\propto t^{-5/6}$}
\label{fig2}
\end{figure}

As shown by the fit in Fig.~\ref{fig2}, the late-time decay follows the power-law behavior
$$\psi\propto t^{-5/6},$$
while the oscillation amplitude is modulated by a sinusoidal term proportional to $\propto\sin{(\mu t)}$. These findings are consistent with previous analyses of massive field perturbations \cite{Koyama:2000hj,  Konoplya:2023fmh, Konoplya:2006gq}.

\subsubsection{Non-Schwarzschild black holes}
\begin{figure*}[hbt!]
    \centering
    \includegraphics[width=\textwidth]{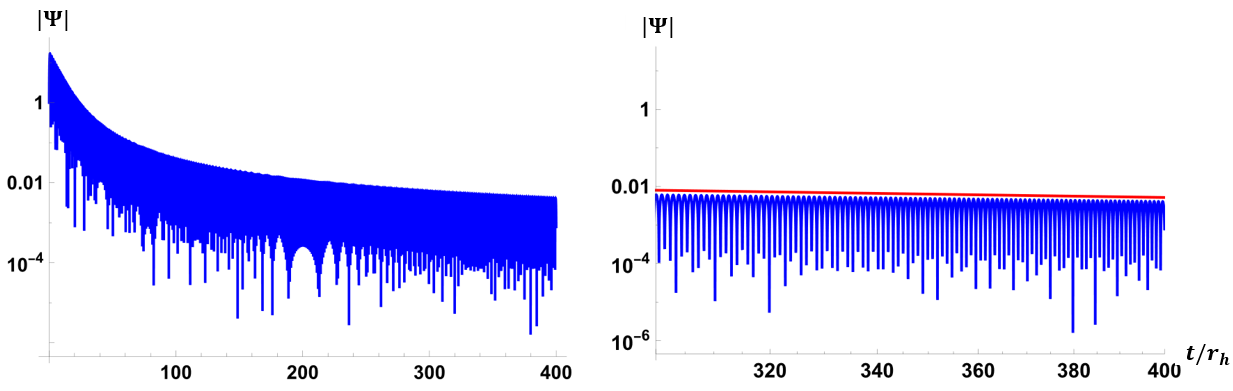}
    \caption{Left panel: Semi Logarithmic plot of time evolution of massive radial perturbation ($l=0$) for the Non-Schwarzschild black hole with $p={\mu} r_h=1$.\\ Right panel: Logarithmic plot of asymptotic tail of the left panel perturbation with a fit given by the law $\propto t^{-5/6}$. }
\label{fig3}
\end{figure*}

Considering the alternative branch of black hole solutions, it is necessary to use the full expression for the potential given by Eq.~\eqref{eq:masterEq_monopole}. Additionally, the metric function must be approximated using the method described in Sec.~\ref{approx}, which is meaningful only within the range of stability of the parameter $p$ \cite{Kokkotas:2017zwt}. We numerically integrated the wave-like equations to get the time evolution of perturbations of this branch of black holes, with representative results shown in Fig.~\ref{fig3}.
In this case, the decay following the ringdown phase is more pronounced, exhibiting a drop of approximately four orders of magnitude. However, the late-time tail again follows the power-law behavior
$$\psi\propto t^{-5/6},$$
confirming the universality of this decay law also for certain massive perturbations.

\subsection{l=2}
For higher multipole numbers, the analysis becomes more involved due to the structure of the tensor harmonics. Nevertheless, the computations were carried out in \cite{Antoniou:2024jku}, where the equations governing gravitational perturbations were derived. The resulting system consists of the classical, massless component $h_1$ and the two massive components, one vectorial $F_1$ and one tensorial $F_2$, which are dynamically coupled. As in the monopole case, we organize our analysis based on the choice of background solution. 

\subsubsection{Schwarzschild black holes}
In the Schwarzschild case, the equations reduce to Eqs.~(\ref{eq:linear-Ricciflat-eom-f}) and (\ref{eq:linear-Ricciflat-eom-g}). The corresponding coupled system, expressed in terms of the tortoise coordinate, takes the form:
\begin{equation} \label{system}
    -\frac{d^2}{dt^2}\boldsymbol{\Psi}+\frac{d^2}{dr_*^2}\boldsymbol{\Psi}+\boldsymbol{P}\frac{d}{dr_*}\boldsymbol{\Psi}+\boldsymbol{V}\Psi=0,
\end{equation}
where $\boldsymbol{\Psi}=\big(\Psi^{(1)},\Psi^{(2)},\Psi^{(3)}\big)\equiv (H_1,Q,Z)$, and $H_1\equiv A h_1/r , \; Q\equiv F_1 A , \; Z\equiv F_2/r.$ The matrices $P$ and $V$ are the potential matrices of the following form:
\begin{equation} \label{matrix}
    \boldsymbol{P}=
    \begin{pmatrix}
        0 & 0 & {P}_{13}\\
        0 & 0 & 0\\
        0 & 0 & 0
    \end{pmatrix}\, , \;
    \boldsymbol{V}=
    \begin{pmatrix}
        {V}_{11} & {V}_{12} & {V}_{13}\\
        0 & {V}_{22} & {V}_{23}\\
        0 & {V}_{32} & {V}_{33}
    \end{pmatrix} \, ,
\end{equation}
whose elements are given by the following relations:
\begin{align}
    P_{13} =& -2 \mu^2 A(r) \nonumber\\
    V_{12}=& -\frac{2 \mu^2 A(r)}{r} \nonumber\\
    V_{13} =& \frac{2 \mu^2 A(r)^2}{r} - 2 \mu^2 A(r) A'(r) \nonumber\\
    V_{11} =& -\frac{A(r)}{r^2} \left( \Lambda - 3 r A'(r) \right)\nonumber\\
    V_{22}=&-A\mu^2+\big[-2 A^2 \left(-r A'+10 A+2 \left(\Lambda -1\right)\right)\nonumber\\
    &+A r^2 A'^2+A r (-2 A r A''-r A'^2+6 A A'\nonumber)\big]\\
    &(4r^2 A)^{-1}\,,\nonumber\\[2mm]
    V_{23}=&-r^{-2}\left(\Lambda -2\right) A\left(2 A-r A'\right)\,,\label{grav_pot_Schw}\\[2mm]
    V_{32}=&-2r^{-2}A\, ,\nonumber\\[2mm]
    V_{33}=&-A\mu^2+\big[r (-A A')-A (r A'\nonumber\\
    &+2 (\mu ^2 r^2+\Lambda-2))(2 r^2)^{-1}\big]\,. \nonumber
\end{align} 
It can be observed that, in this case, the massive modes form a self-contained decoupled subsystem. However, to capture the full dynamics of the perturbations, it is necessary also to include the first equation of the system, which couples all three modes.

Employing the time-domain method described in Appendix~\ref{appendix}, we analyze the evolution of the perturbations, focusing on stability and extracting relevant physical information. As the first step, we consider, as an illustrative example,  the case $p=1$, analyzing the ringdown phase (Fig.~\ref{fig4}). Using the Prony's method (see Appendix~\ref{appendix}), we extract the dominant frequency from the first component of $\boldsymbol{\Psi}$, whose perturbations contain the massless mode along with subdominant contributions from the massive modes. The resulting frequency is $\omega=0.732-0.177i$ deviating from the corresponding General Relativity value by less than $2\%$.

\begin{figure}[ht]
    \centering
    \includegraphics[width=\columnwidth]{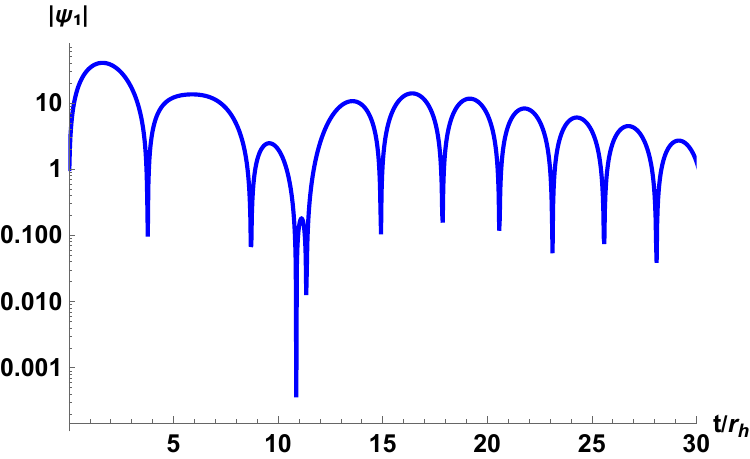}
    \caption{Semi Logarithmic plot of time evolution of perturbation ($l=2$) for Schwarzschild black holes with $p={\mu} r_h=1$. We extract the behavior for the component $\psi_1\propto h_1$ which has contributions of the classical massless perturbations with the peculiar massive perturbations (see eqs.~\eqref{matrix},~\eqref{system}). We extract fundamental frequency with Prony's method and compare with massless Schwarzschild GR value.}
\label{fig4}
\end{figure}

\begin{figure}[ht]
    \centering
    \includegraphics[width=\columnwidth]{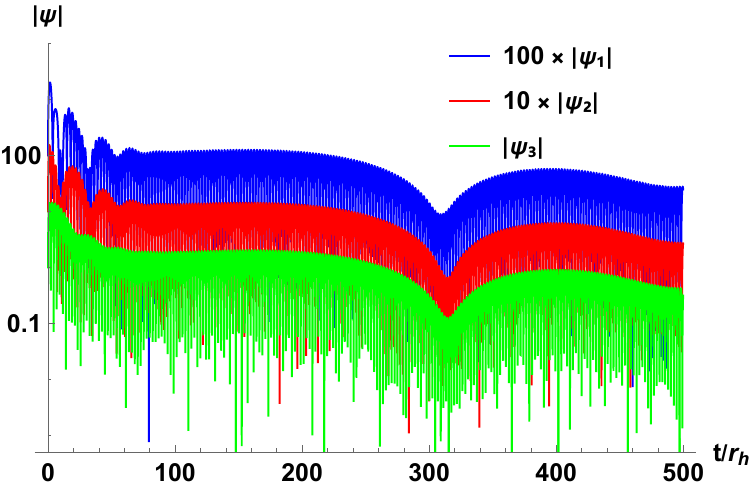}
    \caption{Semi Logarithmic plot of time evolution of perturbation ($l=2$) for Schwarzschild black holes with $p={\mu} r_h=1$. We extract the behavior for the component $\psi_1$, which has contributions both from massless and massive perturbations, and for $\psi_2,\psi_3$ which have only massive perturbations. The time evolution of all components shows a similar behavior, indicating that the massive modes significantly influence the overall dynamics.}
\label{figSchw}
\end{figure}

As in the previous section, we now turn to the study of the asymptotic tail. However, for small values of \( p \), the behavior is illustrated in Fig.~\ref{figSchw}, 

where we observe the presence of echoes in the intermediate tail for all the components. These echoes persist over an extended period, preventing a clear identification of subsequent decay times. Moreover, the peculiar structure of the signal in this regime hinders the extraction of a simple analytical decay law.

\begin{figure}[ht]
    \centering
    \includegraphics[width=\columnwidth]{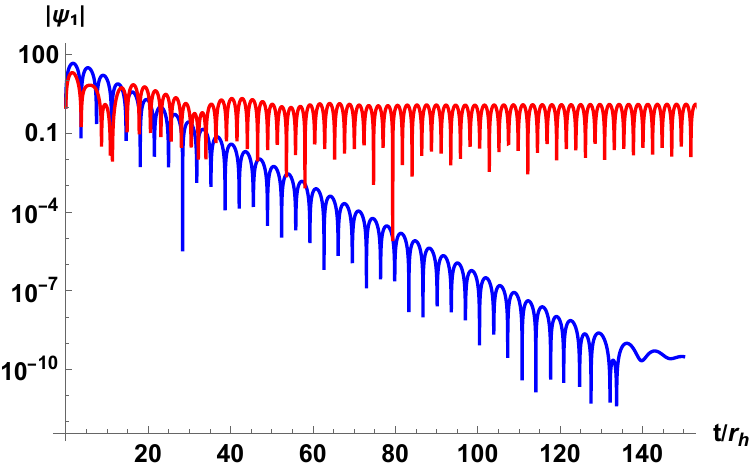}
    \caption{Semi Logarithmic plot of time evolution of perturbation ($l=2$) for Schwarzschild black holes in GR in blue and with $p={\mu} r_h=1$ in EW gravity in red. For the latter we have extracted the behavior for the component $\psi_1$ which has contributions both from massless and massive perturbations.}
\label{figGR-EW}
\end{figure}

Nevertheless, the initial ringdown phase can still be meaningfully compared to that of classical massless gravitational perturbations in General Relativity, as shown in Fig.~\ref{figGR-EW}. 

The presence of massive modes significantly modifies the decay profile, resulting in a much smoother signal. The amplitude differs by several orders of magnitude compared to the massless case, underscoring the substantial influence of the massive sector on the early-time dynamics.

\begin{figure}[ht]
    \centering
    \includegraphics[width=\columnwidth]{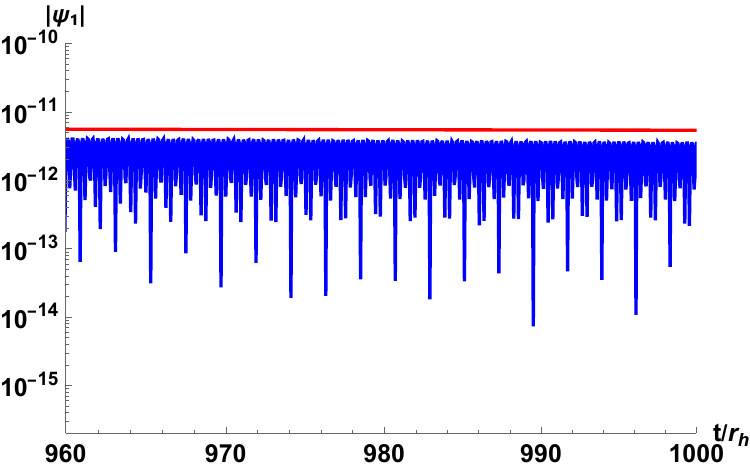}
    \caption{Logarithmic plot of the asymptotic tail of perturbation ($l=2$) for Schwarzschild black holes with $p=r_h\mu=10$ and in red the fit by the law $\propto t^{-5/6}$}
\label{fig5}
\end{figure}

Furthermore, for the Schwarzschild branch, there is no upper bound on the parameter $p$, and increasing its value leads to an earlier onset of the asymptotic tail. Characterizing this regime, we extract the late-time decay law by varying the mass parameter $\mu$ and considering higher multipole numbers $l$, in order to verify the independence of the asymptotic behavior from these parameters. The resulting trend is shown in Fig.~\ref{fig5} and is:
\begin{equation}
    \psi\propto t^ {-5/6}.
\end{equation}

We found that the late-time decay law observed in all other asymptotic tails studied, both in this work and in some other theoretical frameworks \cite{Koyama:2000hj,  Konoplya:2023fmh, Konoplya:2006gq}., consistently follows the same time dependence, further supporting the universality of this asymptotic behavior of massive fields in the evolution of perturbations.

\subsubsection{Non-Schwarzschild black holes}

For the Non-Schwarzschild branch of black holes, the structure of the equations can be reduced to the previous one 
\begin{equation}
    -\frac{d^2}{dt^2}\boldsymbol{\Psi}+\frac{d^2}{dr_*^2}\boldsymbol{\Psi}+\boldsymbol{P}\frac{d}{dr_*}\boldsymbol{\Psi}+\boldsymbol{V}\Psi=0,
\end{equation}
where $\boldsymbol{\Psi}=(\Psi^{(1)},\Psi^{(2)},\Psi^{(3)})$ are obtained connecting these functions to the original components as:
\[
\begin{array}{rcl}
h_1(r) &=& \dfrac{\mu^2 r^2}{\sqrt{A(r) B(r)}} \Psi_1(r),\\
F_1(r) &=& \dfrac{1}{\sqrt{A(r) B(r)}} \Biggl( \Psi_2(r) + \dfrac{\Psi_1(r)}{2 r A(r)} \dfrac{d}{dr} \left(A(r) B(r) r^2\right) \\&&- \Psi_1(r) \Biggr),\\
F_2(r) &=& \Psi_3(r) + \Psi_1(r) \dfrac{r B(r)}{\sqrt{A(r) B(r)}},
\end{array}
\]
and the matrices take the following form
\begin{equation}
    \boldsymbol{P}=
    \begin{pmatrix}
        0 & 0 & {P}_{13}\\
        0 & 0 & 0\\
        0 & 0 & 0
    \end{pmatrix}\, , \;
    \boldsymbol{V}=
    \begin{pmatrix}
        {V}_{11} & {V}_{12} & {V}_{13}\\
        {V}_{21} & {V}_{22} & {V}_{23}\\
        {V}_{31} & 0 & {V}_{33}
    \end{pmatrix} \, ,
\label{eq:axial_system}
\end{equation}
with the following matrix elements:
\begin{align} 
    V_{22}(r) =& (4 r B A' - A (4 + l + l^2 + r^2 \mu^2 \nonumber\\
    &- 4 r B') (r^{-2}) \nonumber\\
    V_{21}(r) =& (-2 + l + l^2) (3 r B A' +\nonumber\\
    & A (-4 + 3 r B'))(2 r^2)^{-1} \nonumber\\
    V_{23}(r) =& -\frac{(-2 + l + l^2) \sqrt{AB} (2 A - r A')}{r^3} \nonumber\\
    V_{31}(r) =& 3 r B^2 A'^2 + 2 A B A'(-2 + 3 r B') \nonumber\\
    & + A^2 (4 r \mu^2 - 12 r \mu^2 B- 4 B' \nonumber\\
    & + 3 r B'^2)(4 \sqrt{AB}r (2 A - r A'))^{-1} \nonumber\\
    V_{33}(r) =& \frac{r B A' - A (l + l^2 + r^2 \mu^2 - r B')}{r^2} \label{grav_pot_NonSchw} \\
    P_{13}(r) =& -\frac{2 A}{r^2} \nonumber\\
    V_{12}(r) =& -\frac{2 A}{r^2} \nonumber\\
    V_{13}(r) =& \frac{2 \sqrt{AB}r (2 A - r A')}{r^3} \nonumber\\
    V_{11}(r) =& -\frac{(-2 + l + l^2) A}{r^2}.\nonumber
\end{align} 

\begin{figure*}
    \centering
    \includegraphics[width=\textwidth]{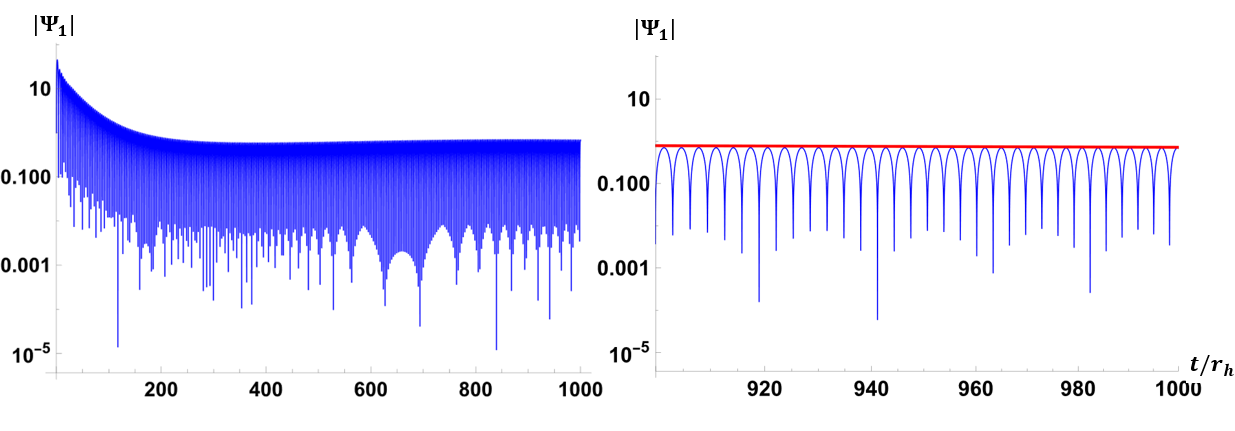}
    \caption{Left panel: Semi Logarithmic plot of time evolution of perturbation ($l=2$) for Non-Schwarzschild black holes with $p={\mu} r_h=1$. We extract the behavior for the component $\psi_1$ which has contributions both from massless and massive perturbations.\\
    Right panel: Logarithmic plot of the asymptotic tail of the left panel perturbation with a fit given by the law $\propto t^{-5/6}$.}
\label{fig6}
\end{figure*}

Since the system in Eq.~\eqref{eq:axial_system} is fully coupled through off-diagonal terms in the potential, perturbations in any single mode will dynamically source the others. As a result, both massless and massive components contribute to the evolution and to the spectrum of the perturbations. We proceeded with the analysis of the time evolution for this branch and observed that the characteristic tails associated with the massive modes emerge early in the signal,  and significantly affect the decay, preventing, thereby,  precise extraction of quasi-normal frequencies. At later times (see Fig.~\ref{fig6}), we confirmed the stability of the black holes within the range of the parameter $p$ where $\ell=0$ perturbations are stable. 

Furthermore, we analyzed the asymptotic tail (right panel of Fig.~\ref{fig6}), and in this case, we were able to study the asymptotically late-time behavior already for $p=1$. We found that the non-Schwarzschild branch also exhibits a power-law decay of the form
\begin{equation}
    \psi \propto t^{-5/6},
\end{equation}
accompanied by oscillatory behavior of the amplitude. In general, the oscillation phase is described by a nontrivial function. However, in the particular case \( p = 1 \), the signal follows the simple form \( \propto \sin{(\mu t + \phi)} \). More generally, the behavior can be expressed as
\begin{equation}
    \psi \propto t^{-5/6} \sin{(F(\mu) t + \phi)},
\end{equation}
where \( F \) is an unknown function that is difficult to determine via numerical fitting.


\section{Eikonal Regime and the Null Geodesic Correspondence}\label{sec:eikonal}

In this section, we investigate the behavior of quasinormal modes in the eikonal regime (i.e., for large multipole number \( \ell \))~\cite{Bolokhov:2025uxz,Konoplya:2023moy}, where in classical general relativity the centrifugal-like term dominates the effective potential. In this limit, the WKB method usually becomes applicable for computing quasinormal frequencies.

In~\cite{Cardoso:2008bp}, a correspondence was proposed between the real and imaginary parts of quasinormal frequencies in the eikonal regime and the properties of unstable circular null geodesics. Specifically, for any stationary, spherically symmetric, and asymptotically flat black hole spacetime, the quasinormal frequencies were conjectured to follow:
\begin{equation} \label{eikonal}
    \omega = \Omega_c\left(l+\tfrac{1}{2}\right) - i\left(n + \tfrac{1}{2}\right)|\lambda|+\Order{l}^{-1},
\end{equation}
where \( \Omega_c \) is the angular velocity and \( \lambda \) the Lyapunov exponent of the corresponding circular null geodesic.

However, it was later shown in~\cite{Konoplya:2017wot,Konoplya:2022gjp} that this correspondence holds reliably only for test fields in stationary, spherically symmetric black hole backgrounds. For gravitational or other non-minimally coupled perturbations — especially in theories with higher curvature corrections such as various Einstein–Gauss–Bonnet models~\cite{Konoplya:2019hml,Konoplya:2020bxa} — the correspondence may break down. This breakdown occurs when the effective potential is no longer a single positive-definite peak that monotonically decays toward both the event horizon and spatial infinity (or a de Sitter horizon).

As a counterexample, in quadratic gravity, the effective potential for gravitational perturbations develops a deep negative gap near the event horizon, which can even diverge. In such cases, quasinormal modes clearly cannot be described by the WKB formula. Indeed, examining the expressions for the potential in eqs.~\eqref{grav_pot_Schw} for the Schwarzschild solution and eqs.~\eqref{grav_pot_NonSchw} for the non-Schwarzschild black hole, we see that in the limit \( l \to \infty \), both cases contain off-diagonal terms such as
\begin{equation} 
V_{23} = - l^2 A(r) \frac{2A - rA'}{r^2}+\Order{l},
\end{equation}
which introduce a negative gap structure inconsistent with the classical potential barrier picture.

To test whether the null geodesic correspondence holds in quadratic gravity, we computed quasinormal frequencies using Prony’s method (as WKB cannot be applied due to the coupled equations and the nature of the potential). For the Schwarzschild background with \( \ell = 100 \), we obtained the fundamental mode:
\begin{equation} \label{frequency}
    \omega r_h = 80.62 - 0.33i.
\end{equation}
By contrast, applying the geodesic-based formula~\eqref{eikonal} yields:
\begin{equation} 
\omega r_h = \dfrac{201-i}{3\sqrt{3}}\approx 38.68 - 0.19i.
\end{equation} 
The discrepancy between these values confirms that the correspondence fails in this regime due to the different structure of the effective potential. The geodesic approach misses essential features of the spectrum, especially in the presence of higher curvature corrections.

It is important to note, however, that for test fields in quadratic gravity, the correspondence still holds, as shown in~\cite{Zinhailo:2018ska}.

\section{Discussion and Conclusion}\label{sec:conclusion}

In this work, we have presented a detailed analysis of the time evolution of gravitational perturbations in black hole spacetimes within Einstein–Weyl gravity. By considering both Schwarzschild and non-Schwarzschild solutions, we have examined the dynamical behavior of monopole and higher multipole perturbations across a broad range of the parameter \( p \), including the near-extremal regime.

Our results reveal several noteworthy features. First, we confirmed the existence of long-lived, oscillatory tails in the time-domain profiles of massive perturbations, following the power-law decay \( \psi \propto t^{-5/6} \). This behavior is universal across both black hole branches and persists for different multipole numbers. Second, we demonstrated that the eikonal correspondence between quasinormal modes and null geodesics breaks down for gravitational perturbations in quadratic gravity due to the nontrivial structure of the effective potential. The WKB approximation fails in this case, and precise frequency extraction requires full numerical treatment.

For small values of the parameter \( p = \mu r_h \), we observed intermediate-time echo structures that obscure the extraction of quasinormal frequencies, further highlighting the complexity of perturbation dynamics in this theory. Nonetheless, our results suggest stability of the solutions within the particular parameter range, consistent with earlier linear stability analyses \cite{Held:2022abx,Antoniou:2024jku,Bonanno:2024fcv}.

While our study provides a comprehensive picture of the dominant quasinormal modes and late-time tails, important questions remain for future work, such as the role of overtones. Accurate extraction of these subdominant modes requires refined numerical methods beyond the scope of the present work. Additionally, the structure of the effective potential near the event horizon may significantly influence the early-time ringdown phase \cite{Konoplya:2022pbc}. Understanding how the horizon geometry affects the excitation and decay of perturbations could shed light on fundamental aspects of black hole dynamics in higher-derivative gravity.

A first comparison with observational data might lead one to consider the bounds placed by LIGO on the graviton mass \cite{LIGOScientific:2020tif}. However, it is important to note that such constraints are derived from modifications to the dispersion relation of gravitational waves, under the assumption that the graviton is massive — that is, within the framework of a massive gravity theory where the primary propagating degree of freedom is itself massive. In contrast, Einstein-Weyl gravity is not a massive gravity theory in this sense: the main propagator remains the massless graviton, while an additional massive spin-2 degree of freedom appears, contributing to the gravitational interaction through a Yukawa-type potential. As a result, while the short range dynamics have contributions of massive terms, the long-range dynamics are still governed by the standard massless graviton.

Thus, the bounds for the graviton mass for the Yukawa potential from the Solar System, galaxy clusters, weak lensing and other observations of the large scale structures \cite{deRham:2016nuf} do not allow one to find an estimation for $\mu$ because the corresponding contribution in \eqref{pert} is exponentially suppressed if we assume that the  graviton's Compton length is much smaller than the Solar System size.

Consequently, the most relevant bounds on the mass parameter $\mu$ come from experimental tests of deviations from the inverse-square law of gravity with Yukawa-like contributions \cite{PhysRevLett.98.021101}. These include both astrophysical observations -- such as gravitational redshift measurements on solar system \cite{Accioly:2015fka} and white dwarfs, or Pound-Rebka-Snider experiments and laboratory experiments, like torsion balance setups probing short-range interactions \cite{Giacchini:2016nta}. The absence of observed deviations from classical gravity translates into an upper bound on the strength of the Yukawa correction, i.e. on the coupling parameter (having a constrain value of $\alpha<10^{81}$ for redshift experiment and $\alpha<10^{61}$ with torsion test \cite{Giacchini:2016nta}), which, in turn, implies a lower bound on the mass $\mu$, $\mu>10^{-13} eV/c^2$ and $\mu>10^{-3} eV/c^2$, respectively. This is physically intuitive: a larger mass leads to a shorter interaction range, making the corresponding degree of freedom harder to excite in scattering processes and less detectable in gravitational measurements. Thus, no significant upper bound can be meaningfully set on $\mu$, since in the limit $\mu\xrightarrow{}\infty$ the theory smoothly reduces to standard Einstein-Hilbert gravity (see eq.\eqref{eq:action} — a feature that is also evident when analyzing the perturbed field equations and the time evolution of the perturbations if one goes to larger $\mu$.

\acknowledgments

We would like to acknowledge Alfio Bonanno, G. Antoniou, L. Gualtieri and P. Pani.  The work of A. S.  was supported by University of Catania and INFN Sezione di Catania. A.S. is grateful for the hospitality of the Institute of Physics at Silesian University where this work was carried out.
\appendix 

\section{Numerical method} \label{appendix}
Starting from the general second-order equation governing our perturbations, we have
\begin{equation}
    -\frac{d^2}{dt^2}\boldsymbol{\Psi}+\frac{d^2}{dr_*^2}\boldsymbol{\Psi}+\boldsymbol{P}\frac{d}{dr_*}\boldsymbol{\Psi}+\boldsymbol{V}\Psi=0.
\end{equation}
Our goal is to compute the time evolution of $\psi$. To do so, we discretize the wave function and the potential matrices on a grid as follows:
$$
\begin{array}{rclcl}
\Psi(r_*, t) &=& \Psi(j \Delta r_*, i \Delta t) &=& \Psi_{j,i}, \\
    V(r(r_*)) &=& V(j \Delta r_*) &=& V_j, \\
    P(r(r_*)) &=& P(j \Delta r_*) &=& P_j.
\end{array}
$$
We then apply our discretization method to approximate the derivatives using the following scheme \cite{Zhu_2014}:
\begin{align}\notag
   &
    - \frac{\Psi_{j,i+1} - 2\Psi_{j,i} + \Psi_{j,i-1}}{\Delta t^2} + \frac{\Psi_{j+1,i} - 2\Psi_{j,i} + \Psi_{j-1,i}}{\Delta r_*^2}
    \\
   & +\frac{P_j}{2\Delta r_*} (\Psi_{j+1,i} - \Psi_{j-1,i})+ V_j \Psi_{j,i} +\Order{\Delta t^2,\Delta r_*^2} = 0.
\end{align}
To solve this equation, we first need to express the potentials as functions of the tortoise coordinate $r_*$. This requires determining the dependence of the radial coordinate $r$ on $r_*$, which we obtain by solving the following differential equation numerically:
\begin{equation}
    \frac{dr}{dr_*}=\sqrt{A(r)B(r)}.
\end{equation}
Once this transformation is obtained and substituted into the potential, we can proceed solving the equation. Imposing the initial condition as followed
\begin{equation} \label{gauss}
    \psi_1=\psi_2=\psi_3=\psi(r_*,0)=e^{-\frac{(r_*)^2}{2\sigma}},
\end{equation}
we choose a Gaussian distribution for all components, centering it at $r_*=0$ to ensure it is located near the peak of the potential. This is also the point where we monitor the time evolution of the perturbation.
To extract the frequencies values from time evolution we used Prony's method \cite{Konoplya:2011qq}, assuming to fit the data by a combination of complex damped exponential
\begin{equation} \label{prony}
    \phi(t)\simeq \sum_{i=1}^p C_i e^{-i\omega_i t}
\end{equation}
and at the end, you can obtain the frequencies as \cite{Konoplya:2011qq,pozrikidis2011introduction,Berti:2007dg},
\begin{equation}
    \omega_j=\frac{i}{h}\log z_j
\end{equation}
where $h$ is the time sampling interval, and $z_j$ are the roots of polynomials obtained from the exponential in eq.~\eqref{prony}.

\bibliography{whbibesterna}

\end{document}